\documentclass[aps,twocolumn,pra,tightenlines]{revtex4}
\usepackage{amsmath,amsthm,amsfonts,amssymb,amsbsy,mathrsfs}
\usepackage{units,times,float,graphicx,bm,comment}
\usepackage[colorlinks,citecolor=blue,linkcolor=blue,hyperindex,dvipdfm]{hyperref}
\begin{document}
\title{Margolus-Levitin speed limit across quantum to classical regimes based on trace distance}
\author{Shao-xiong Wu$^1$\footnote{sxwu@nuc.edu.cn}}
\author{Chang-shui Yu$^2$\footnote{ycs@dlut.edu.cn}}
\affiliation{$^1$ School of Science, North University of China,
Taiyuan 030051, China\\
$^2$ School of Physics, Dalian University of Technology, Dalian 116024, China}
\date{\today }

\begin{abstract}
The classical version of Mandelstam-Tamm speed limit based on the Wigner function in phase space is reported by B. Shanahan et al. [Phys. Rev. Lett. 120, 070401 (2018)]. In this paper, the Margolus-Levitin speed limit across the quantum-to-classical transition  is given in phase space based on the trace distance. The Margolus-Levitin speed limit is set by the Schatten $L_1$ norm of the generator of time dependent evolution for both the quantum and classical domains. As an example, the time-dependent harmonic oscillator is considered to illustrate the result. \\[-6pt]


\end{abstract}

\maketitle

\section{Introduction}
The notion of quantum speed limit (QSL) was proposed by Mandelstam and Tamm \cite{Mandelstam45} in 1945. The Mandelstam-Tamm (MT) bound of QSL, which is defined by the variance of the energy $\frac{\pi\hbar}{2\Delta E}$,  can be considered as the extension of the Heisenberg time-energy relation.  Later, Margolus and Levitin  \cite{Margolus98,Levitin09} found another bound form of QSL (ML bound), which is based on the mean energy $\frac{\pi\hbar}{2(\langle E\rangle-E_0)}$.  The ML bound can be considered as the transition probability amplitude between two orthogonal quantum states $\langle\psi_0\vert\psi_{\tau}\rangle$. In order to get the tight bound, the unified QSL is defined by the larger one between the MT and ML bound, i.e., $\tau_{\text{qsl}}=\max\left\{\frac{\pi\hbar}{2\Delta E},\frac{\pi\hbar}{2(\langle E\rangle-E_0)}\right\}$.

The investigation of QSL can deepen the understanding of time-energy uncertainty relation, the quantum evolution, the quantum control \cite{Caneva09}, even the information of black hole \cite{Bekenstein81}. In particular, the QSL  has been investigated in different methods for the closed systems \cite{Fleming73,Bhattacharyya83,Anandan90,Pati91,Vaidman92,Brody03,Jones10,Campaioli18}, and extended to the open systems recently \cite{Taddei13,Campo13,Deffner13,Xu14,Zhang14,Sun15,Jing16,Pires16,Wu15,Wu18,Cai17,Zhang16,
Yu18,He16,Liu19,Teittinen19,Funo19,Hou17,Xu19,Zhang18,Mondal16,Sun19,Haseli19,Deffnernjp,Xu18,
Pintos19,Campbell17,Xu14cpl,Feng19,Wang20}. One can also see the comprehensive review article \cite{Deffner17}. According to Bohr's correspondence principle, the effect of the reduced Planck's constant $\hbar$ will vanish gradually when the system transitions to the classical world from quantum scale. So, it will cause both the ML and the MT bounds of QSL to become zero,
\begin{align}
\lim_{\hbar\rightarrow0}\tau_{\text{qsl}}=0.
\end{align}
Considering the quantum speed limit originated from the Heisenberg time-energy uncertainty relationship,  people are usually made to believe that the quantum speed limit is the unique phenomenon of quantum mechanics. However, the speed limit for classical dynamics, unrelated with the quantum/classical nature, were reported by B. Shanahan et al. \cite{Shanahan18} and M. Okuyama et al. \cite{Okuyama18}, independently. Using the fidelity and employing the  Cauchy-Schwarz inequality, B. Shanahan et al. obtained a MT  bound of speed limit based on the quasiprobability distributions in the Wigner function, which can describe the transition across the quantum to classical regime in phase space. From the quantum-to-classical domain, the speed of evolution is determined by the Schatten $L_2$ norm of Moyal bracket for the Hamiltonian, the Wigner function and the Bhattacharyya coefficient, respectively. A natural question is whether there also exists a ML bound for the classical speed limit.

In this paper, we derive the Margolus-Levitin bound for speed limit of classical system in phase space based on trace distance. The ML bound for the speed limit across the quantum-to-classical transition can be obtained through the triangle inequality for integral, and it is governed by the Schatten $L_1$ norm of time dependent evolution generator for both the quantum and classical regimes. As an example, the time-dependent harmonic oscillator is given to illustrate
the result. The distinction between the Margolus-Levitin bound in this paper and the Mandelstam-Tamm bound in Ref. \cite{Shanahan18} are discussed. This paper is organized as follows. In Sec. II, we give the definition of Margolus-Levitin quantum speed limit in phase space. In Sec. III, the ML semiclassical speed limit is defined. In Sec. IV, the ML classical speed limit is obtained, and an example is given. The discussion and the conclusion are given in the end.

\section{The  Margolus-Levitin  quantum speed limit in phase space}

Motivated by Ref. \cite{Shanahan18}, we will derive the ML bound of speed limit in phase space representation. The Wigner function of one-dimensional system $\rho_t$ is defined as \cite{Hillery84,Zachos05}
\begin{align}
W_t(q,p)=\frac{1}{\pi\hbar}\int\langle q-y\vert{\rho}_t\vert q+y\rangle e^{-2ipy/\hbar}dy,
\end{align}
where $q,p$ are the generalized coordinate and momentum respectively, and $\langle q\vert{\rho}_t\vert q'\rangle$ is the density matrix in the coordinate representation.  In the following paper, we will consider the unitary dynamics of pure state under the time dependent Hamiltonian ${H}$, and the ``distance" measure between the initial state and final state is chosen as the trace distance, which is given by
\begin{align}
\mathcal{T}(\rho_t,\rho_0)=\text{Tr}|\rho_t-\rho_0|=\|\rho_t-\rho_0\|_1.
\end{align}
Turning into the Wigner phase space, the distance $\mathcal{T}(W_t,W_0)$ between the pure initial state with Wigner function $W_0$ and the time-dependent final state with Wigner function $W_t$ is \cite{Deffnernjp}
\begin{align}
\mathcal{T}(W_t,W_0)=\int d^2\Gamma |W_t-W_0|,
\end{align}
where $d^2\Gamma=2\pi\hbar dqdp$.

In order to consider the time-dependent  change rate of the trace distance, the motion of Wigner function $W_t$ can be expressed by the Moyal bracket \cite{Zachos05}
\begin{align}
\frac{\partial W_t}{\partial t}=\{\{H,W_t\}\}=\frac{1}{i\hbar}(H_{qp}\star W_t-W_t\star H_{qp}),\label{qsl:moyal}
\end{align}
where $\{\{H,W_t\}\}$ is the Moyal bracket and $\star$-product means the Moyal product $
H_{qp}\star W_t=H_{qp}\text{exp}\Big(\frac{i\hbar}{2}\overleftarrow{\partial_q}\overrightarrow{\partial_p}
-\frac{i\hbar}{2}\overleftarrow{\partial_p}\overrightarrow{\partial_q}\Big)W_t(q,p)
$ with Weyl ordered Hamiltonian operator $H_{qp}=\int dx\langle q-x/2\vert H\vert q+x/2\rangle\exp(ipx/\hbar)$. Similar to Refs. \cite{Deffnernjp,Cai17}, the rate of change for the trace distance $\mathcal{T}(W_t,W_0)$ can be derived as
\begin{align}
\dot{\mathcal{T}}(W_t,W_0)=\int d^2\Gamma \frac{W_t-W_0}{|W_t-W_0|}\{\{H,W_t\}\}.\label{Eq7}
\end{align}
Using the property of integral (or the triangle inequality for integral), it can lead to the following inequality:
\begin{align}
\dot{\mathcal{T}}(W_t,W_0)\leq|\dot{\mathcal{T}}(W_t,W_0)|\leq\int d^2\Gamma |\{\{H,W_t\}\}|.
\end{align}
It should be note that the value of $(W_t-W_0)/|W_t-W_0|$ is $\pm1$. And, it is easy to obtain that
\begin{widetext}
\begin{align}
\int d^2\Gamma |\{\{H,W_t\}\}|&=\int d^2\Gamma \left|\frac{\partial W_t}{\partial t}\right|\notag\\
&=\int 2\pi dqdp\langle q-y\vert(\vert\dot{\rho}_t\vert)\vert q+y\rangle\langle q+y'\vert\mathbb{I}\vert q-y'\rangle e^{2ip(y'-y)/\hbar}dydy'\notag\\
&=2\int dqdy\langle q-y\vert(\vert\dot{\rho}_t\vert)\vert q+y\rangle\langle q+y\vert\mathbb{I}\vert q-y\rangle\notag\\
&=2\int\frac{dXdY}{2}\langle X\vert(\vert\dot{\rho}_t\vert)\vert Y\rangle\langle Y\vert X\rangle\notag\\
&=\text{Tr}|\dot{\rho}_t|,\label{tuidao}
\end{align}
\end{widetext}
where $\mathbb{I}$ means the identical density matrix.
Substituting  the Heisenberg equation $i\hbar{\partial_t{\rho}_t}=[H,\rho_t]$ into Eq. (\ref{tuidao}), we can get the following inequality:
\begin{align}
\text{Tr}\vert\dot{\rho}_t\vert&=\frac{1}{\hbar}\text{Tr}\vert[\rho_t,H]\vert\notag\\
&\leq\frac{1}{\hbar}\text{Tr}\vert\rho_tH\vert+\frac{1}{\hbar}\text{Tr}\vert H\rho_t\vert.\label{qsl:bds}
\end{align}
In the second line, the triangle inequality for trace norm is used. For the normalized pure state, we can obtain that
\begin{align}
\text{Tr}\vert\rho_tH\vert=\text{Tr}\sqrt{H\vert\psi_t\rangle\langle\psi_t\vert H}=\langle H\rangle,
\end{align}
where $\langle H\rangle$ is the mean energy. Without loss of generality, following Ref. \cite{Margolus98}, assuming that the system has discrete spectrums, and the energy eigenvalues $\{E_n\}$ associated with the eigenstates $\{\vert E_n\rangle\}$ are in ascending order. When choosing the energy of ground state properly and assuming that the value of ground state $E_0$ is zero, we can have
\begin{align}
\text{Tr}\vert\rho_tH\vert=\langle H\rangle-E_0\label{eq12}
\end{align}

Combining the above Eqs. (\ref{Eq7}-\ref{eq12}), we arrive at
\begin{align}
\vert\dot{\mathcal{T}}(W_t,W_0)\vert\leq\int d^2\Gamma\vert\{\{H,W_t\}\}\vert\leq\frac{2(\langle H\rangle-E_0)}{\hbar}.\label{eq13}
\end{align}
Integrating Eq. (\ref{eq13}) over time from $0$ to $\tau$, we can obtain that
\begin{align}
\tau\geq\frac{\hbar\mathcal{T}(W_{\tau},W_0)}{2E_{\tau}}
\geq\frac{\mathcal{T}(W_{\tau},W_0)}{\langle v_{\text{qsl}}\rangle},\label{qsl}
\end{align}
where
\begin{align}
E_{\tau}=(1/\tau)\int_0^{\tau}dt(\langle H\rangle-E_0)
\end{align}
is the time-averaged energy, and
\begin{align}
\langle v_{\text{qsl}}\rangle=(1/{\tau})\int_0^{\tau}dt\|\{\{H,W_t\}\}\|_1
\end{align}
is defined as the time-averaged velocity in the quantum information processing. The Schatten $L_1$ norm of Moyal bracket
\begin{align}
\|\{\{H,W_t\}\}\|_1=\int d^2\Gamma\vert\{\{H,W_t\}\}\vert=v_{\text{qsl}}
\end{align}
can be considered as the instantaneous velocity of quantum state evolution in phase space. Eq. (\ref{qsl}) has the same form as the original Margolus-Levitin speed limit $\frac{\hbar\pi}{2(\langle E\rangle-E_0)}$ and is related to the value of time-averaged energy, so it can be considered as Margolus-Levitin quantum speed limit bound in phase space
\begin{align}
\tau\geq\tau_{\text{qsl}}=\frac{\mathcal{T}(W_{\tau},W_0)}{(1/{\tau})\int_0^{\tau}dt\|\{\{H,W_t\}\}\|_1},\label{qsl:defination}
\end{align}
which is governed by the Schatten $L_1$ norm of Moyal bracket.

\section{The  Margolus-Levitin semiclassical speed limit}

The Moyal bracket can be expanded through Taylor expansion in term of $\hbar$, and will be reduced to the Poisson bracket when ignoring the higher-order terms. It can be expressed as follows \cite{Zachos05}
\begin{align}
\{\{H,W_t\}\}=\{H,W_t\}+\mathcal{O}(\hbar^2).
\end{align}
The Poisson bracket is defined as functions of the partial derivatives of generalized canonical coordinate $q$ and momentum $p$, i.e.,
\begin{align}
\{f,H\}=\frac{\partial H}{\partial p}\frac{\partial f}{\partial q}-\frac{\partial H}{\partial q}\frac{\partial f}{\partial p},
\end{align}
and it governs the dynamics of classical mechanics.

Due to the Wigner function containing the reduced Planck's constant $\hbar$,  the semiclassical speed limit (SSL) can also be given based on the Poisson bracket as
\begin{align}
\tau\geq\tau_{\text{ssl}}&=\frac{\mathcal{T}(W_{\tau},W_0)}{\langle v_{\text{ssl}}\rangle},\label{ssl:ml}
\end{align}
where,
\begin{align}\langle v_{\text{ssl}}\rangle=(1/{\tau})\int_0^{\tau}dt\|\{H,W_t\}\|_1
\end{align}
is the time-averaged velocity of system evolution governed by the Poisson bracket. The upper bound of the instantaneous evolution velocity is defined by the absolute value for the change rate of Wigner function averaged over the phase space, i.e.,
\begin{align}
v_{\text{ssl}}&=\int d^2\Gamma \vert\partial_tW_t\vert=\int d^2\Gamma \vert\{H,W_t\}\vert\notag\\
&=\|\{H,W_t\}\|_1.
\end{align}
In order to derive the formula (\ref{ssl:ml}), one only need to replace the Moyal bracket $\{\{H,W_t\}\}$ by the Poisson bracket $\{H,W_t\}$. Because the Wigner function $W_t$ contains the reduced Planck's constant $\hbar$, it is appropriate to say that Eq. (\ref{ssl:ml}) is a Margolus-Levitin semiclassical speed limit (ML-SSL) bound. Independent of the quantum approach, Eq. (\ref{ssl:ml}) can also been derived through the classical mechanics. The Hamilton's equation of motion in classical mechanics is
\begin{align}
\frac{\partial W_t}{\partial t}=\{H,W_t\},
\end{align}
so the semiclassical speed limit can be arrived straightforwardly
\begin{align}
\tau\geq\tau_{\text{ssl}}=\frac{\mathcal{T}(W_{\tau},W_0)}{(1/\tau)\int_0^{\tau}dt\| \hat{L}W_t\|_1}.\label{ssl:ml2}
\end{align}
The Liouvillian $i\hat{L}W_t=-\{H,W_t\}$ is introduced in Eq. (\ref{ssl:ml2}). Similar to Eq. (\ref{qsl:defination}) in the quantum case, the ML-SSL is also set by the Schatten $L_1$ norm of Liouvillian $\hat{L}$. It should be noticed that the expression ML-SSL in phase space is dependent on the Winger function, and still includes the reduced Planck's constant $\hbar$.

\section{The Margolus-Levitin classical speed limit}
Now, we will derive the speed limit for the classical evolution. According to the operational dynamic modeling \cite{Bondar12,Bondar13}, the relationship for the evolution of the dynamical average values between the classical phase-space probability density $\varrho_t(q,p)$ and the Wigner function $W_t(q,p)$ can be connected through the Ehrenfest theorems, which is given by
\begin{align}
\varrho_t(q,p)=2\pi\hbar W_t(q,p)^2.
\end{align}
For a pure state $\vert\psi\rangle$, the normalized condition meets $
2\pi\hbar\int dqdpW_t(q,p)^2=\int dqdp\varrho_t(q,p)=1$. Similar to the original seminal work of Margolus-Levitin quantum speed limit bound derived from the transition probability amplitude between two orthogonal states $\langle\psi_0\vert\psi_t\rangle$, the overlap between states $\varrho_0$ and $\varrho_t$ can be re-expressed by the Bhattacharyya coefficient \cite{Bhattacharyya46}
\begin{align}
B(\varrho_t,\varrho_0)=\int dqdp\sqrt{\varrho_0\varrho_t}.
\end{align}
Reminiscing the Hellinger distance $H(\varrho_t,\varrho_0)^2=\frac{1}{2}\text{Tr}[(\sqrt{\varrho_0}-\sqrt{\varrho_t})^2]$, which can be used to measure the quantum correlation \cite{Wu14,Chang13,Girolami13}, it is easy to verify that $B(\varrho_t,\varrho_0)=1-H(\varrho_t,\varrho_0)^2.$

The trace distance between the initial and final states can be expressed as
\begin{align}
\mathcal{T}(\varrho_t,\varrho_0)=\int dqdp\vert\sqrt{\varrho_t}-\sqrt{\varrho_0}\vert,
\end{align}
and the classical speed limit can be derived straightforwardly as
\begin{align}
\tau\geq\tau_{\text{csl}}&=\frac{\mathcal{T}(\varrho_{\tau},\varrho_0)}{\langle v_{\text{csl}}\rangle},\label{csl:ml}
\end{align}
where,
\begin{align}
\langle v_{\text{csl}}\rangle=(1/{\tau})\int_0^{\tau}dt\|\{H,\sqrt{\varrho_t}\}\|_1
\end{align}
is the  time-averaged velocity of classical system dynamics. The instantaneous classical evolution velocity is defined as the Schatten $L_1$ norm of $\{H,\sqrt{\varrho_t}\}$
\begin{align}
v_{\text{csl}}&=\|\{H,\sqrt{\varrho_t}\}\|_1=\int dqdp\left\vert\{H,\sqrt{\varrho_t}\}\right\vert\notag\\
&=\int dqdp\vert\partial_t\sqrt{\varrho_t}\vert.
\end{align}
Eq. (\ref{csl:ml}) is a classical version of Margolus-Levitin speed limit bound for system evolution, and it can be obtained using only the classical approach. The classical Liouville operator satisfies $\partial_t\varrho_t+i\hat{L}\varrho_t=0$, so the derivative of trace distance is
\begin{align}
\dot{\mathcal{T}}(\varrho_t,\varrho_0)&\leq\vert \dot{\mathcal{T}}(\varrho_t,\varrho_0)\vert\notag\\
&=\int dqdp\vert \hat{L}\sqrt{\varrho_t}\vert=\|\hat{L}\sqrt{\varrho_t}\|_1.\label{csl:zyjl}
\end{align}
Integrating Eq. (\ref{csl:zyjl}) over time from $0$ to $\tau$, we can obtain that
\begin{align}
\tau\geq\tau_{\text{csl}}=\frac{\mathcal{T}(\varrho_{\tau},\varrho_0)}{(1/\tau)\int_0^{\tau}dt\|\hat{L}\sqrt{\varrho_t}\|_1}.\label{scl}
\end{align}
This is the main result of this paper, and it is a classical Margolus-Levitin version bound of speed limit for classical dynamics.

Similar to Ref. \cite{Shanahan18}, we would like to consider the time-dependent harmonic oscillator as an example to illustrate the above results. The time-dependent harmonic oscillator can be applied to the control protocols \cite{Campbell17,Chen10}, the quantum thermal machines \cite{Campo14}, etc. The Hamiltonian is given by
\begin{align}
\hat{H}=\frac{{p}^2}{2m}+\frac{1}{2}m\omega(t)^2{q}^2.
\end{align}
In the quantum case, the Wigner function of state under the modulation of trapping frequency $\omega(t)$ is
\begin{align}
W_n(q,p;t)=&W_n(Q,P;0)\notag\\
=&\frac{(-1)^n}{\pi\hbar}e^{-(2/\hbar\omega_0)[P^2/(2m)+(1/2)m\omega_0^2Q^2]}\notag\\
&\times L_n\Big[\frac{4}{\hbar\omega_0}\Big( \frac{P^2}{2m}+\frac{1}{2}m\omega_0^2Q^2\Big)\Big],
\end{align}
where $L_n[x]$ is the Laguerre polynomials and $Q=q/b$, $P=bp-mq\dot{b}$ are the pairs of canonically conjugated  variables, respectively. The time-dependent factor $b(t)$ is governed by the Ermakov equation $\ddot{b}+\omega(t)^2b=\omega_0^2/b^3$, where the boundary conditions are $b(0)=1$ and $\dot{b}(0)=0$ \cite{Chen10,Chen10prl}.

Suppose that the initial ground state of harmonic oscillator with $n=0$, and $W_0(q,p,t)\geq0$ is a smooth Gaussian distribution for all $0\leq t\leq\tau$. The classical distribution is chosen as the Gaussian form $\varrho_0(q,p)=\exp(-q^2/\sigma_q^2-p^2/\sigma_p^2)/(\pi\sigma_q\sigma_p)$ and $\sigma_q=x_0/\sqrt{2}$, $\sigma_p=\hbar/(x_0\sqrt{2})$. The state $\varrho_t(q,p)$ can be evaluated by $\varrho_t(q,p)=\varrho_0(Q,P)$, and the Bhattacharyya coefficient can be calculated analytically as
\begin{align}
B(\varrho_0,\varrho_t)=2\Big[\frac{(1+b^2)^2}{b^2}+\Big(\frac{m\sigma_q\dot{b}}{\sigma_p}\Big)^2\Big]^{-1/2}.\label{csl:b}
\end{align}
The upper Margolus-Levitin bound of evolution instantaneous velocity in phase space is
\begin{align}
v_{\text{csl}}=\| H,\sqrt{\varrho_t}\|_1=\sqrt{\frac{\sigma_q}{\pi\sigma_p}}4m\sigma_q\vert\ddot{b}(t)\vert.\label{v:ml}
\end{align}

\begin{figure}
\begin{center}
   \includegraphics[width=1\columnwidth]{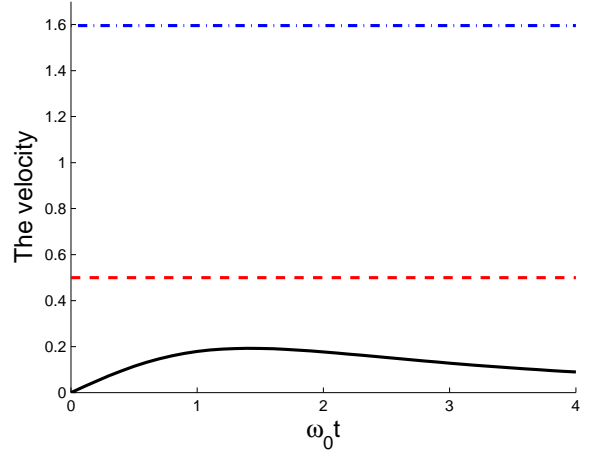}
\caption{ Classical speed limit of evolution in phase space. The black solid line is the absolute value of the Bhattacharyya coefficient derivative $\vert\dot{B}(\varrho_0,\varrho_t)\vert$; The red dashed line is the Mandelstam-Tamm upper bound of speed $v_{\Gamma}^{\text{CSL}}$ reported in Ref. \cite{Shanahan18}; The blue dot-dashed line is the  Margolus-Levitin upper bound of instantaneous speed $v_{\text{csl}}$ in Eq. (\ref{v:ml}). The unit of time is set by $\omega_0^{-1}$.}\label{fig}
\end{center}
\end{figure}

The driven Hamiltonian is assumed to be a constant for $t>0$, and the frequency of trapping turns off suddenly at $t=0$. One can find that $b(t)=\sqrt{1+\omega_0^2t^2}$ and $\ddot{b}(t)=\omega_0^2$. In Fig. \ref{fig}, we show the Margolus-Levitin velocity $v_{\text{csl}}$ in phase space, the Mandelstam-Tamm velocity $v_{\Gamma}^{\text{CSL}}$ reported in Ref. \cite{Shanahan18} and the absolute value of the Bhattacharyya coefficient derivative $\vert\dot{B}(\varrho_0,\varrho_t)\vert$ during the evolution, respectively. What needs illustration is that the Margolus-Levitin speed limit (\ref{scl}) can not be compared with the Mandelstam-Tamm speed limit in Ref. \cite{Shanahan18}, because the original ``distance"  measures of them are different.

\section{Discussion and conclusion}
In this paper, we obtained the Margolus-Levitin speed limit across the quantum to classical regime based on the trace distance. However, the trace distance is not a bona fide measure to derive the Mandelstam-Tamm speed limit in phase space, because the Cauchy-Schwarz inequality for $n\times n$-dimensional matrix, i.e., $\vert\text{Tr}A\vert\leq\sqrt{n}\sqrt{\text{Tr}(A^{\dag}A)}$, can not be extended to infinite dimension system directly. In addition, to the best of our knowledge, since the von Neumann inequality can only be applied to matrices, the fidelity or the overlap between states, which can lead to the Mandelstam-Tamm speed limit in phase space, is not appropriate to be employed as the measure of Margolus-Levitin speed limit in phase space. The unified speed limits across the quantum to classical worlds are still absent. Among the possible unified distance measures, the Hilbert-Schmidt norm distance may not a good candidate due to the non-contractivity \cite{Ozawa00,Piani12}. Since, the Wigner function is fundamental in the field of quantum optics, the investigation of speed limit across the quantum to classical regimes in phase space and discussing its properties (such as the achievable, the tightness) can deepen the understanding about the quantum control, quantum dynamical property, statistical behavior of classical systems, etc. The nature of speed limit still deserves our further investigation.

In summary, the speed limit is not the unique phenomenon of quantum system. In this paper, utilizing the trace distance and triangle inequality for integral, the Margolus-Levitin speed limit bound across the quantum-to-classical transition is obtained in phase space. We find that the Margolus-Levitin bound of speed limit is governed by the Schatten $L_1$ norm of the dynamical generator. As an example, the time-dependent harmonic oscillator is given to illustrate the results.

\section*{Acknowledgments}
Project supported by the National Natural Science Foundation of China (Grant No. 11775040), Scientific and Technological Innovation Programs of Higher Education Institutions in Shanxi (Grant No. 2019L0527), Fundamental Research Fund for the Central Universities (Grant No. DUT18LK45).

\end{document}